\documentclass[%
twocolumn,
bibnotes,
superscriptaddress,
 amsmath,amssymb,
 aps,
 prb
]{revtex4-1}

\usepackage{graphicx}% Include figure files
\usepackage{dcolumn}% Align table columns on decimal point
\usepackage{bm}% bold math
\usepackage{blindtext}
\usepackage{multirow}
\usepackage{color}
\usepackage{float}

\begin{document}

%\preprint{}

\title{Topological Fermi-arc surface resonances in bcc iron}


\author{Daniel Gos\'albez-Mart\'inez}
\affiliation{Institute of Physics, Ecole Polytechnique F\'ed\'erale de Lausanne (EPFL), CH-1015 Lausanne, Switzerland}
\affiliation{National Centre for Computational Design and Discovery of Novel Materials MARVEL, Ecole Polytechnique F\'ed\'erale de Lausanne (EPFL), CH-1015 Lausanne, Switzerland}

\author{Gabriel Aut\`es}
\affiliation{Institute of Physics, Ecole Polytechnique F\'ed\'erale de Lausanne (EPFL), CH-1015 Lausanne, Switzerland}
\affiliation{National Centre for Computational Design and Discovery of Novel Materials MARVEL, Ecole Polytechnique F\'ed\'erale de Lausanne (EPFL), CH-1015 Lausanne, Switzerland}

\author{Oleg V. Yazyev}
\email{E-mail: oleg.yazyev@epfl.ch}
\affiliation{Institute of Physics, Ecole Polytechnique F\'ed\'erale de Lausanne (EPFL), CH-1015 Lausanne, Switzerland}
\affiliation{National Centre for Computational Design and Discovery of Novel Materials MARVEL, Ecole Polytechnique F\'ed\'erale de Lausanne (EPFL), CH-1015 Lausanne, Switzerland}

%\thanks{}


\date{\today}

\begin{abstract}
The topological classification of matter has been extended to include semimetallic phases characterized by the presence of topologically protected band degeneracies. In Weyl semimetals, the foundational gapless topological phase, chiral degeneracies are isolated near the Fermi level and give rise to the Fermi-arc surface states. However, it is now recognized that chiral degeneracies are ubiquitous in the band structures of systems with broken spatial inversion ($\mathcal{P}$) or time-reversal ($\mathcal{T}$) symmetry. This leads to a broadly defined notion of topological metals, which implies the presence of disconnected Fermi surface sheets characterized by non-zero Chern numbers inherited from the enclosed chiral degeneracies. Here, we address the possibility of experimentally observing surface-related signatures of chiral degeneracies in metals. As a representative system we choose bcc iron, a well-studied archetypal ferromagnetic metal with two nontrivial electron pockets. We find that the (110) surface presents arc-like resonances attached to the topologically nontrivial electron pockets. These Fermi-arc resonances are due to two different chiral degeneracies, a type-I elementary Weyl point and a type-II
composite (Chern numbers $\pm 2$) Weyl point, located at slightly different energies close to the Fermi level. We further show that these surface resonances can be controlled by changing the orientation of magnetization, eventually being eliminated following a topological phase transition. Our study thus shows that the intricate Fermi-arc features can be observed in materials as simple as ferromagnetic iron, and are possibly very common in polar and magnetic materials broadly speaking. Our study also provides methodological guidelines to identifying Fermi-arc surface states and resonances, establishing their topological origin and designing control protocols.
\end{abstract}

%\keywords{}

\maketitle

\section{\label{sec:intro} Introduction}

Band degeneracies in condensed matter have attracted considerable interest due to their relation to the topological character of metallic systems \cite{weyl_murakami,PhysRevB.83.205101}, while quasiparticles in proximity to these band degeneracies emulate different types of relativistic particles \cite{PhysRevB.83.205101,PhysRevX.5.011029,DS_1} and even extend beyond the conceived high-energy schemes \cite{Burkov11,Soluyanov2015,Wang2016,Bradlynaaf5037}. The energy dispersion around a band crossing, intimately related to the symmetries of the system, determines the nature of the quasiparticles. Such fermion quasiparticles are topologically protected and become especially relevant when the band degeneracies are close to the Fermi level giving rise to the notion of topological semimetals \cite{weyl_review}. 

The most generic case of a band degeneracy occurs in systems where $\mathcal{PT}$ symmetry, the combination of inversion ($\mathcal{P}$) and time-reversal ($\mathcal{T}$) symmetries, is broken. In this situation, the Kramers degeneracy is lifted, and band crossings between two bands may be present at isolated points in the three dimensional (3D) Brillouin zone (BZ). The band dispersion around the crossing point is generally linear\cite{Herring_1} and is described by the $2\times2$ generalized Weyl Hamiltonian $H(\textbf{q})= \sum_{i,j} q_i \nu_{ij} \sigma_j$, where $\textbf{q}=\textbf{k}-\textbf{k}_0$ is the position in $k$-space relative to the degeneracy point $\textbf{k}_0$, $\nu_{ij}$ are real coefficients that determine the chirality of the Weyl fermion and $\sigma_j$ are the Pauli matrices including the identity matrix. 
Such Weyl node degeneracies cannot be easily gapped by small perturbations. The Berry curvature $\Omega(\textbf{k})$ presents a singularity when two bands are degenerate. Therefore, a Weyl node acts as a monopole of the Berry curvature in momentum space, carrying a chiral charge $\chi$ that is determined by the total flux of $\Omega(\textbf{k})$ in units of 2$\pi$ through a surface enclosing the degeneracy. The only possibility to open a gap at a Weyl point is by annihilating it with another Weyl point of opposite chiral charge. 

When $\mathcal{PT}$ symmetry is preserved the energy bands are doubly degenerate, thus any band crossing involves a 4-fold band degeneracy. In this case, the fermion quasiparticles around the band touching point are described by the $4\times4$ Dirac Hamiltonian. Unlike their Weyl counterparts, Dirac points are not topologically protected since their chiral charge is zero. One can think of a Dirac point as two superimposed Weyl nodes of opposite chiral charge stabilized by some point group symmetry \cite{DS_1,DS_2,DS_3}. 

Both Weyl and Dirac fermions can be divided into two different types according to the morphology of the energy dispersion around the band touching point. Type-I Weyl fermions present the standard conical energy dispersion. Such band degeneracies create closed Fermi surfaces which collapse into a single point when the Fermi level is at the energy of the Weyl point. Type-II Weyl fermions have a tilted conical energy dispersion. In contrast to the type I , these degeneracies produce open electron and hole Fermi surfaces which undergo a Lifshitz transition at the energy of the Weyl point \cite{Soluyanov2015}.
Band degeneracies carrying chiral charges of $\chi = \pm 2$ and $\pm 3$ are also possible. These composite band degeneracies occurring at $C_4$ or $C_6$ rotation symmetry axes are characterized by either quadratic or cubic energy dispersions in the plane perpendicular to the symmetry axis. Such band crossings might be considered as double or triple Weyl points stabilized by the rotation symmetry \cite{multi_weyl,Stepan}. 

This taxonomic classification of quasiparticles can be further extended to include non-conventional relativistic fermions. For example, nonchiral nodal lines 
%(lines of degeneracies) 
can occur in the presence of mirror or non-symmorphic symmetries\cite{PhysRevB.92.081201,1674-1056-25-11-117106}. In these cases, the symmetry reduces the co-dimension of the system allowing the bands to be degenerate along lines instead of points\cite{Burkov11}. Furthermore, 3-, 6- and 8-band crossing with either linear or quadratic band dispersion can appear at high symmetry points under specific space groups. Some of these degeneracies carry a chiral charge and can be considered as the spin-1 and spin-3/2 generalizations of Weyl fermions\cite{Bradlynaaf5037}.

Band degeneracies, due to their topological nature, are associated with novel physical properties. Point nodes present a new type of surface states referred to as the Fermi arcs whose energy dispersion is defined between the projections of two nodes of opposite chirality on the surface Brillouin zone (sBZ) \cite{wan11}. At the Fermi level, a finite Fermi arc contour that starts and ends at the projections of the two chiral nodes is produced. Likewise, lines of degeneracies have been shown to exhibit surface states with a drumhead shape \cite{Liu2014,Liu864,Huang2015,PhysRevLett.116.066802}. Furthermore, new transport properties such as the chiral anomaly\cite{NIELSEN1983389,PhysRevB.88.104412} and novel quantum oscillations due to the presence of Fermi arcs \cite{Potter2014}, are present when charge carriers are described by chiral fermions. Ideally, to observe these properties, the band degeneracies have to be close to the Fermi level, and isolated from the rest of the bands. The search for materials that present isolated band crossings at the Fermi level has been exhaustive and many candidates have been proposed. For further details, we would like to refer the reader to recent review articles on this topic \cite{0953-8984-28-30-303001,doi:10.1146/annurev-conmatphys-031016-025458,weyl_review,doi:10.7566/JPSJ.87.041001,doi:10.1021/acs.chemmater.7b05133}. 

In general, however, it is fair to say that band degeneracies may appear at any energy, and are not necessarily isolated from other bands. In the case of metallic systems, Fermi surface sheets provide closed surfaces on which Chern numbers can be defined (Fermi surface Chern numbers), therefore providing an avenue to their topological classification. In this context, a metal that has at least a pair of Fermi surface sheets with Chern numbers different from zero is considered a topological metal \cite{haldane1}. This definition of topological metal comprises the topological semimetals, which are the limiting case with Fermi surfaces collapsing to single points. 

Charge carriers enclosed by the nontrivial Fermi surface sheets retain their chiral nature. Similar to the chiral nodes, surface projections of nontrivial Fermi sheets with opposite Chern numbers are connected by the Fermi arcs. In this case, the topological surface states are well defined only in the regions of the sBZ where no bulk states are projected. Unfortunately, in many metals with extended Fermi surfaces, most of the sBZ is covered by projected bulk states, and it is likely that the Fermi arc crosses regions of the sBZ where bulk states are present. At surface momenta where both a surface state and bulk states exist, the former hybridizes with the bulk states becoming a surface resonance that has a bulk character with a strong weight on the surface. This resonance presents a finite lifetime on the surface prior to penetrating into the bulk, which depends on the coupling between the surface and bulk states and is manifested in the broadening of the surface state. When interaction between the surface and bulk states is strong, the broadening of the surface resonance is such that the resonance is indistinguishable from the bulk states. This is one of the main difficulties in observing Weyl semimetals experimentally since the Weyl points often coexist with other bands and are not at the Fermi level. The presence of the bulk states also obscures the exact origin of the Fermi arc.

Given the fact that chiral band degeneracies are ubiquitous in metals with broken $\mathcal{PT}$ symmetry, one can anticipate observing resonances originating from these chiral degeneracies. In this work, we search for such topological resonances in a prototypical ferromagnet--body-centered cubic (bcc) iron--by studying its surface electronic structure. As it was previously shown in Ref.~\onlinecite{bccFe}, bcc Fe is a topological metal whose band structure presents a large number of band degeneracies including single and double Weyl nodes as well as non-chiral nodal lines. Furthermore, the Fermi surface of iron consists of several sheets, most of which are topologically trivial except two Fermi surface pockets with Chern numbers $\mathcal{C}=\pm 1$ when the magnetization is oriented along the [001] easy axis. These nontrivial Fermi surfaces pockets, located along the axis parallel to the magnetization direction, enclose a single Weyl point each and have to be connected by the Fermi arcs. Note that other band degeneracies can also produce surface states associated with Fermi surface sheets of zero net chiral charge \cite{haldane1}. Several surface resonances have been observed experimentally in bcc iron for different surface orientations \cite{KIM2001193,0295-5075-59-4-592,PhysRevB.65.184412,PhysRevLett.92.097205,PhysRevB.72.155115,PhysRevLett.103.267203}, although no direct link between these surface resonances and topological band degeneracies has been established. 

If one is able to unambiguously associate a surface resonance with a chiral degeneracy, it would be possible to determine the chiral nature of band degeneracies even away from the Fermi level. It is of particular interest for identifying time-reversal symmetry broken Weyl nodes that remain elusive with rare exceptions \cite{Liu2017,Kuroda2017}. To associate the origin of a surface resonance with the Fermi arc created by two chiral degeneracies, one needs to determine the exact origin in momentum space of the surface resonance, at least at one end of the Fermi arc. Furthermore, ideally, one seeks to be able to manipulate the topological state of the system by external parameters, \emph{i.e.} being able of annihilate the chiral degeneracies in order to observe a change of the surface resonance upon a topological phase transition. In the case of magnetic metals, the orientation of the magnetization naturally provides such a control parameter. In fact, it has been shown experimentally that the Fermi surface of bcc iron changes appreciably upon varying the magnetization orientation \cite{PhysRevX.6.041048}. We will follow this strategy in our work to identify the topological origin of the surfaces resonances in this simple magnetic system.

Below, we investigate a surface resonance at the (110) surface of bcc iron that we shown to originate from topologically nontrivial Fermi surface sheets. The corresponding Fermi arc that hybridizes with the projected bulk states connects the Weyl point enclosed by a nontrivial Fermi surface sheet with a double Weyl point of opposite chiral charge enclosed by one of the the trivial surfaces. Interestingly, this composite Weyl point is tilted in the direction parallel to the magnetization showing that not only linear band degeneracies can be found in both flavors, either type-I and type-II, but also the double Weyl nodes. When the direction of magnetization is tilted away from the easy axis, the composite Weyl node splits into two linear type-II Weyl points. For specific orientations of the magnetization, one of the two Weyl points enters into the nontrivial pocket changing the net chiral charge inside the pocket to zero, and therefore, resulting in a topological phase transition. The studied surface resonance changes its morphology following the  transition, {\it i.e.} it emerges and submerges into a small electron pocket indicating the trivial nature of this pocket. 

The rest of the article is organized as follows. In Section~\ref{sec:topo_metal} we give some basic definitions of berryology, and present a brief reminder on the electronic structure bcc iron and its topological properties. Exceeding details can be found in Ref.~\onlinecite{bccFe} while in this work we focus on the composite Weyl node responsible for the abovementioned topological phase transition. In Section \ref{sec:surface_states}, we describe the surface density of states of the (110) surface of bcc iron and discuss the nontrivial resonance. In order to gain an insight into the origin of the surface resonance, in Section \ref{sec:topo_tran} we address the topological phase transition driven in by the orientation of the magnetization and its effect on the Weyl points and surface states. Finally, main results of our work are summarized in Section~\ref{sec:conclusion}.

\section{\label{sec:topo_metal} Bcc iron as a topological metal}

\begin{figure*}[t]
\centering
  \includegraphics[width=.95\textwidth]{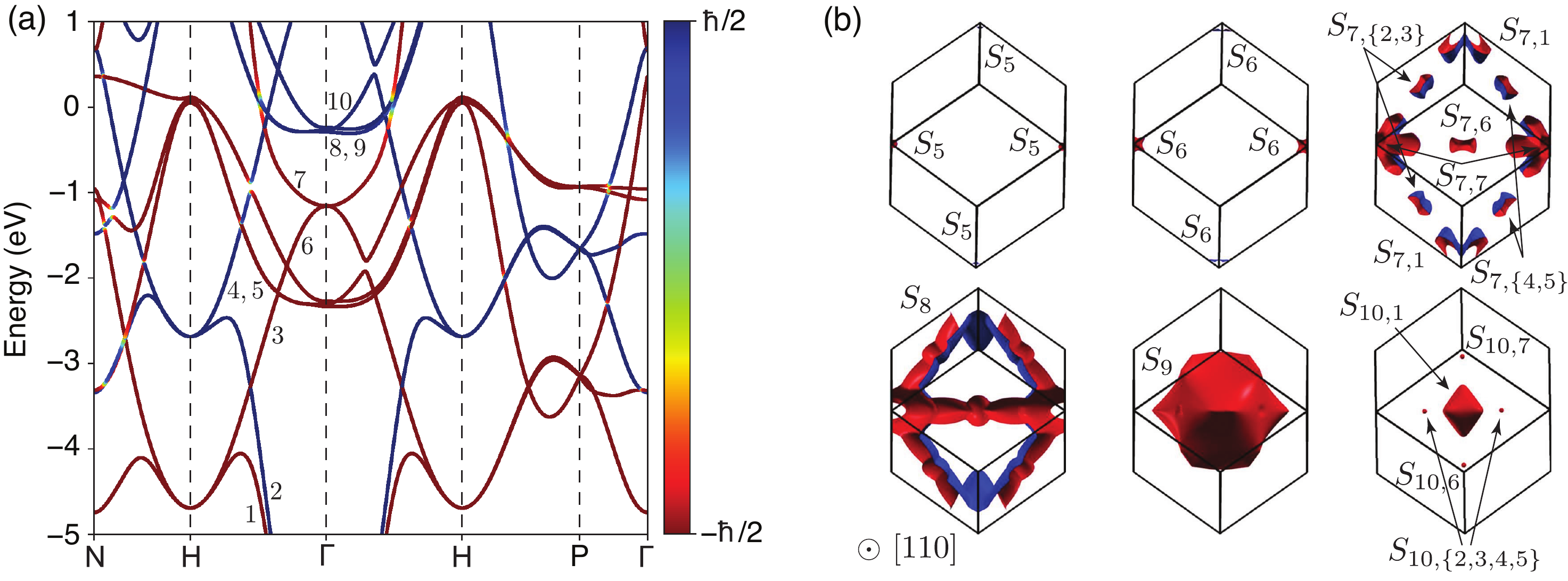} 
  \caption{
(a) Band structure of bcc iron with magnetization along the $[001]$ direction and including SOC. The color indicates the $s_z$ spin polarization, referring to the the spin down (majority) and spin up (minority) polarizations in red and blue, respectively.
(b) Fermi surface sheets of the six bands that cross the Fermi level. The Brillouin zone is viewed along the [110] direction. The color indicates the side of the surface that is in contact with electrons (blue) and holes (red). Each Fermi surface sheet is labeled as $S_{n,i}$, where $n$ indicates the band, and $i$ the individual sheet. Due to the orientation of the BZ some sheets are projected on top of each other, and hence both labels are given.}
\label{fig:1}
\end{figure*}

In order to relate a surface resonance to a chiral degeneracy in a metal, one needs to know the exact distribution of band degeneracies in momentum space. For the case of ferromagnetic iron, the survey of band degeneracies has already been performed in Ref.~\onlinecite{bccFe}. Here, for the sake of completeness, we present a summary of the electronic structure of bcc iron and associated band degeneracies, preceded by the description of computational methodology and relevant definitions. 

\subsection{\label{subsec:topo_metal:0} First-principles methodology}

The electronic structure of bcc iron presented in this section, and in the rest of the article, has been obtained from first principles using the plane-wave pseudopotential method implemented in the \texttt{Quantum-Espresso} package \cite{qe}. Magnetization parallel to the easy axis [001] and the experimental lattice parameter $a=2.87 \mathrm{~\AA}$ were chosen. The exchange-correlation potential was described using the PBE functional within the generalized-gradient approximation~\cite{PBE}. The spin-orbit coupling (SOC) was included through the fully relativistic norm-conserving pseudopotentials fused in Ref.~\onlinecite{bccFe}. A plane-wave kinetic energy cutoff of 120~Ry, a Fermi smearing of 0.02~Ry, and a $16 \times 16 \times 16$ Monkhorst-Pack grid have been employed.

Maximally localized Wannier functions (MLWFs) have been used to interpolate the first-principles band structure to a higher $k$-space resolution and to obtain the Hamiltonian in a localized basis for computing the surface density of states as described below. To construct the MLWFs we used the \texttt{Wannier90} code \cite{wannier90}. Starting from the overlaps of 28 Bloch functions computed on a $8 \times 8 \times  8$ $k$-point mesh, we obtained 18 Wannier functions using as initial spinor projection the $sp^3d^2$ hybrid orbitals and the $d_{xy}$, $d_{xz}$ and $d_{yz}$ atomic orbitals of the iron atom as described in Ref. ~\onlinecite{WYSV06}.

In Sections \ref{sec:surface_states} and \ref{sec:topo_tran} we will use the Hamiltonian in the MLWF basis set to compute the surface density of states for different surface orientations. The surface density of states is obtained from the Green's function of a semi-infinite surface computed by the method described in  Ref.~\onlinecite{surfaceGF1}. The Hamiltonian used for computing the Green's functions was obtained from a principal layer of a supercell containing 6 iron atoms repeated in the [110] direction. 

\subsection{\label{subsec:topo_metal:1} Basic definitions}

The topological nature of a metal is given by the presence of band degeneracies. When two bands $n$ and $n'$ are degenerate in $k$-space, the Berry curvature $\Omega^{n}(\textbf{k})$ of band $n$ presents a singularity, therefore this band degeneracy acts as a monopole of this field. This fact can be explicitly seen defining the Berry curvature as
\begin{equation}
\Omega^{n}_{\alpha \beta}(\textbf{k}) = -2 \mathrm{Im} \sum_{n' \neq n} \frac{ \langle u_{n\textbf{k}} \vert H_{\alpha}(\textbf{k}) \vert u_{n'\textbf{k}} \rangle \langle u_{n'\textbf{k}} \vert H_{\beta}(\textbf{k}) \vert u_{n \textbf{k}} \rangle}{(\epsilon_{n} - \epsilon_{n'})^2},
\end{equation}
where $\vert u_{n \textbf{k}} \rangle$ is the cell-periodic part of the Bloch function $\vert \psi_{n \textbf{k}} \rangle$ of band $n$, and $H_{\alpha}(\textbf{k})=\frac{\partial H(\textbf{k})}{\partial_{k_\alpha}}$ with $H(\textbf{k})=e^{i \textbf{k} \cdot \textbf{r}} H e^{-i \textbf{k} \cdot \textbf{r}}$. 

An equivalent vector definition of the Berry curvature is 
\begin{equation}
\Omega_n{(\textbf{k})} = \nabla  \times \textbf{A}_n(\textbf{k}),
\end{equation}
where $\textbf{A}_n$ is the Berry connection
\begin{equation}
A_n{(\textbf{k})} = i \langle u_{n \textbf{k}} \vert  \nabla_{\textbf{k}} \vert u_{n \textbf{k}} \rangle .
\end{equation}
The relation between these two definitions of Berry curvature is $\Omega_{n,\gamma}(\textbf{k}) = \epsilon_{\alpha \beta \gamma} \Omega_{n,\alpha \beta}(\textbf{k})$, where $\epsilon_{\alpha \beta \gamma}$ is the three-dimensional Levi-Civita symbol. 
According to the Chern theorem, the total flux of the Berry curvature through a closed and orientable surface is quantized in units of 2$\pi$, and defines a topological invariant called the Chern number of a given surface. In metals, the sheets of the Fermi surface define closed surfaces on which it is possible to compute this topological invariant. One can define the Fermi surface Chern number $\mathcal{C}_{n,i}$ of the Fermi surface sheet $i$ of band $n$ as flux integral
\begin{equation}
\mathcal{C}_{n,i}= \frac{1}{2\pi} \int_{S_{n,i}} \Omega_n(\textbf{k}) \cdot \textbf{n} dS.
\end{equation}
This quantity can be defined as the total chiral charge enclosed by the $S_{n,i}$ surface\cite{bccFe}. In the same fashion, the chiral charge of a single chiral node can be obtained by computing the flux through a small surface that encloses just that single point alone. The methodology for evaluating numerically these quantities can be found in Refs.~\onlinecite{doi:10.1143/JPSJ.74.1674,bccFe,PhysRevB.95.075146}.

\subsection{\label{subsec:topo_metal:2} Electronic structure}

The magnetic point group of ferromagnetic bcc iron is $4/mm'm'$ for the magnetization pointing along the easy axis [001]. It contains a $C_4$ rotation symmetry axis parallel to the orientation of the magnetization, a mirror symmetry plane perpendicular to the magnetization, {\it i.e.} the (001) plane, and two antisymmetric mirror reflection planes, (010) and (110). However, these symmetries can be eliminated by changing the direction of the magnetization to an arbitrary orientation. This will later help us in analyzing the topological protection of different types of band degeneracies.

The band structure of ferromagnetic iron, shown in Figure~\ref{fig:1}a, is an ideal playground for illustrating a broad range of band degeneracies that can be found in condensed matter systems. At first glance, we can see that band degeneracies are ubiquitous in this band structure. In the case of ferromagnetic bcc iron, only accidental degeneracies can be found when SOC is included in the description~\cite{PhysRev.172.498,PhysRevB.1.1261}. A systematic study of all the band degeneracies in bcc Fe was reported in Ref.~\onlinecite{bccFe}. Here, we provide a brief overview of the electronic structure and band degeneracies for the following two purposes: first, to illustrate the different types degeneracies that might appear in $\mathcal{PT}$ broken systems, and second, to point out precisely the degeneracies that lead to the appearance of surface resonances.
 
Bcc iron has six bands that cross the Fermi level giving rise to a complex Fermi surface. In Figure \ref{fig:1}b we illustrate the Fermi surface sheets associated with each band. Labels $S_{n,i}$ refer to sheet $i$ of band $n$ following the notation introduced in Ref.~\onlinecite{bccFe}. Most sheets surround the inversion symmetric points $\Gamma$, $\mathrm{H}$ and $\mathrm{N}$, and consequently, their Fermi surface Chern numbers are zero. The only Fermi sheets that can give rise to Chern numbers different from zero are the 6 electron pockets $S_{10,i}$ ($i=2,...,7$). The $S_{10,i}$ electron pockets can be related by symmetry according to the orientation of the magnetization. In the case considered here, for the magnetization along the easy axis [001], the six $S_{10,i}$ pockets are divided into two inequivalent subgroups. The four $S_{10,i}$ $(i=2,...,5)$ pockets in the $k_z=0$ plane are related by the $C_4$ rotation symmetry. The $S_{10,6}$ and $S_{10,7}$ pockets located on the $k_z$ axis are related by the $M_z$ mirror symmetry. In fact, as it was shown in Ref.~\onlinecite{bccFe}, these two pockets along the $\Gamma H$ direction parallel to the magnetization vector enclose Weyl points, and consequently have Fermi surface Chern numbers $\mathcal{C}_{10,6}=-1$ and $\mathcal{C}_{10,7}=+1$. The four electron pockets on the $k_z=0$  mirror symmetry plane are connected to the Fermi surface sheet $S_9$ through a nodal line. Due to the presence of such degeneracy on the surface, strictly speaking, one cannot define the corresponding Chern numbers, although a tiny shift of the magnetization can break the symmetry  detaching the Fermi surface sheets.

\subsection{\label{subsec:topo_metal:3} Band degeneracies}

\begin{figure}[t]
\centering
  \includegraphics[width=.48\textwidth]{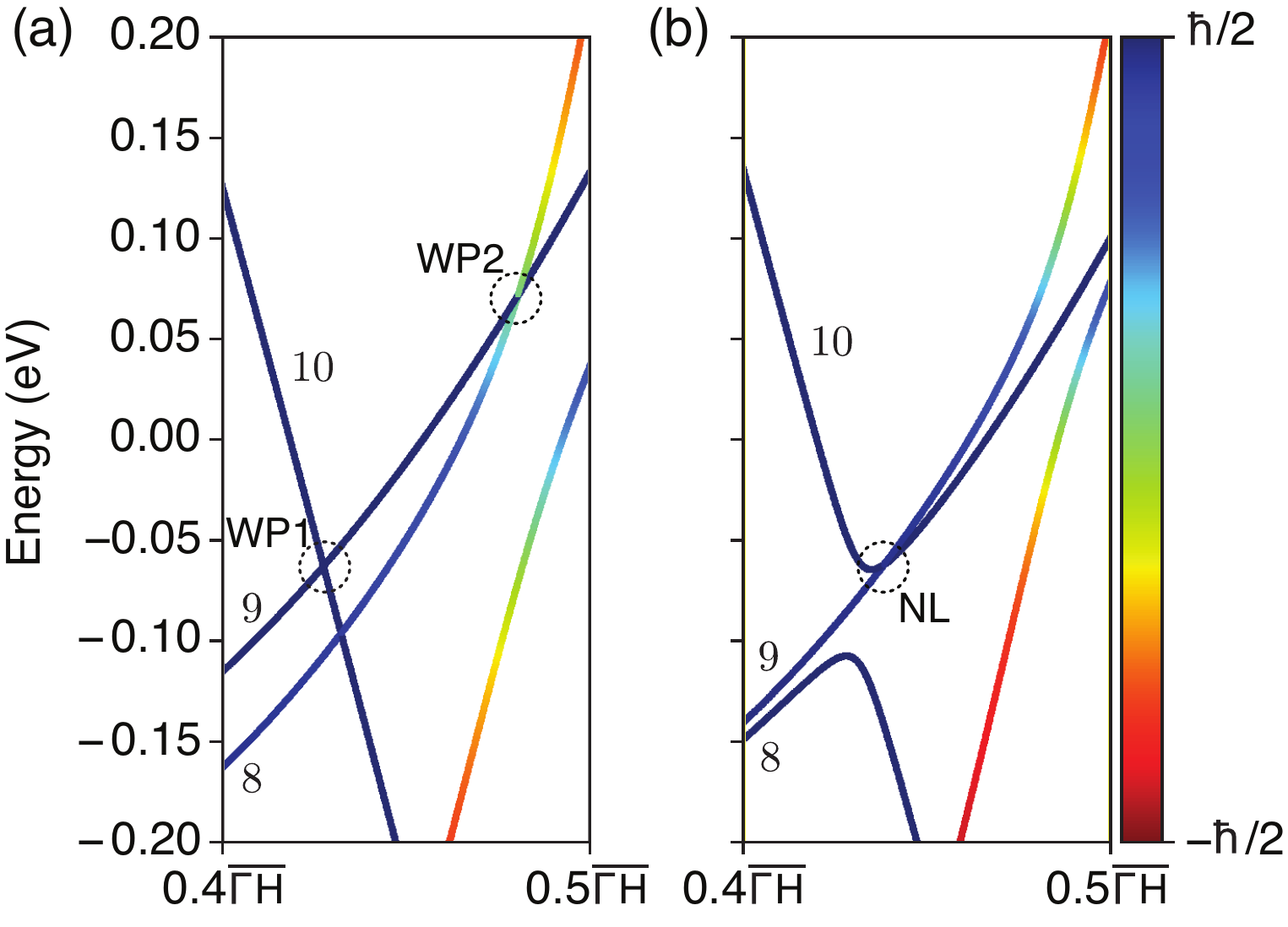} 
  \caption{\label{fig:bands} Details of the band structure of bcc iron along the $\Gamma H$ direction (a) parallel and (b) perpendicular to the magnetization. The color denotes the $s_z$ component of the spin polarization. Elementary (WP1) and double (WP2) Weyl nodes are indicated in panel (a). In panel (b), a band crossing related to a nodal line in the $k_z=0$ mirror symmetry plane is shown.}
\end{figure}

Below, we will focus on the topologically nontrivial Fermi surface sheets $S_{10,i}$ ($i=2,...,7$) and band degeneracies in their neighborhood. Band 10 crosses both bands 9 and 11 at different points in the BZ. However, in a region close to the electron pockets, the only band crossings of band 10 are with band 9 along the two non-equivalent $\Gamma H$ directions as shown in Figure~\ref{fig:bands}. These band crossings illustrate the different types of band degeneracies that are present in bcc iron.

\begin{figure*}[ht!]
\includegraphics[width=.78\textwidth]{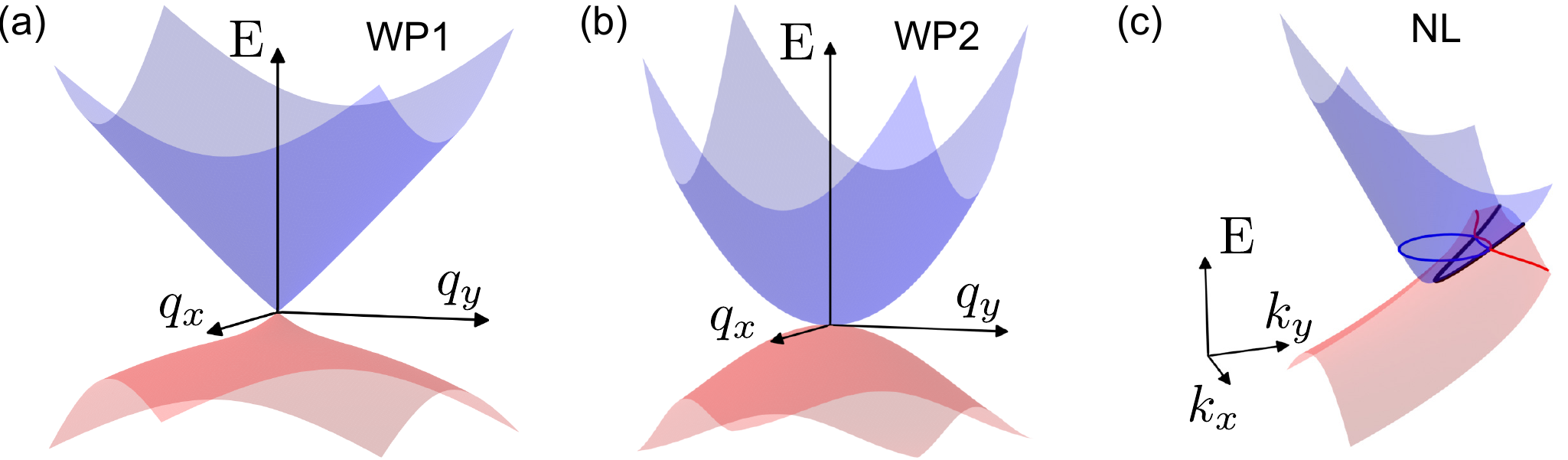} 
\caption{\label{fig:WP} Band dispersion around (a) the elementary Weyl point, (b) the double Weyl point and (c) the nodal line that are present close to electron pockets $S_{10,i}$ ($i=2,...,7$).   
In panel (c), we show the Fermi contour lines as red and blue lines to illustrate how the nodal line (in black) pierces the two Fermi surfaces.}
\end{figure*}

\paragraph*{Elementary Weyl nodes.} The band degeneracy enclosed by electron pockets $S_{10,6}$ and $S_{10,7}$ is shown in Figure~\ref{fig:bands} and will be referred to as WP1. This degeneracy at $k_z \approx 0.42 (2\pi/a)$ and 63~meV below the Fermi level is a typical example of Weyl point of chiral charge $\chi = \pm 1$. It is characterized by a linear band dispersion in all directions, as one can see in Figures~\ref{fig:bands}(a) and \ref{fig:WP}(a). 
Since this point and its mirror symmetric counterpart are the only degeneracies enclosed by the $S_{10,6}$ and $S_{10,7}$ electron pockets, the resulting Fermi surface Chern numbers of these pockets are $- 1$ and $+1$, respectively. 

\paragraph*{Double Weyl nodes.} An example of composite Weyl node can be found along the same axis at $k_z \approx 0.48 (2\pi /a)$ and 71~meV above the Fermi level (Fig.~\ref{fig:bands}(b)). Close to this band touching point that we denoted WP2, the energy dispersion is linear along the $k_z$ direction, that is the $C_4$ rotation symmetry axis. In the perpendicular plane, however, the dispersion is quadratic as one can see in Figure~\ref{fig:WP}(b). This type of band degeneracy can be considered as two elementary Weyl points brought together by the $C_4$ rotation symmetry, as explained in Refs.~\onlinecite{multi_weyl,Stepan}, and consequently, such composite Weyl points carry a chiral charge $\chi = \pm 2$. 

The WP2 band degeneracy presents a tilted energy dispersion along the $k_z$ axis with both involved bands having positive Fermi velocities (Fig. \ref{fig:bands}(b)). Hence, we classify this band degeneracy as the type-II composite Weyl node. We note that such classification into two types according to the morphology of the Fermi surface around the band degeneracy has been developed exclusively for the elementary Weyl nodes with linear dispersion of bands. Generalization of such classification to composite Weyl nodes is not straightforward, and to the best of our knowledge has not been done. In the case of the composite Weyl nodes, the Fermi surfaces are described by algebraic surfaces of $4^{th}$ or $6^{th}$ order whose classification is not complete.

We can illustrate the complexity of the morphology of composite degeneracies by considering the low-energy Hamiltonian in the neighborhood of a double Weyl node generated by the $C_4$ rotation symmetry. Following the notation of Ref.~\onlinecite{Stepan}, we write the Hamiltonian as
\begin{equation}
H(\textbf{q}) = d(\textbf{q}) \mathbb{I} + f(\textbf{q}) \sigma_{+} + f^{*}(\textbf{q}) \sigma_{-} + g(\textbf{q}) \sigma_z,
\end{equation}
where $\mathbb{I}$ is the identity matrix, $\sigma_{\pm}=\sigma_x \pm i \sigma_y$, with $\sigma$ being the Pauli matrices, and $\textbf{q}=\textbf{k}-\textbf{k}_0$ with $\textbf{k}_0$ being the position of the nodal point. As it was shown in Ref.~\onlinecite{Stepan}, due the $C_4$ rotation symmetry, the momentum dependence of functions $f(\textbf{q})$ and $g(\textbf{q})$ is given by
\begin{eqnarray}
f(\textbf{q}) & = & aq_{+}^{2}+bq_{-}^{2}, \\
g(\textbf{q}) & = & l q_z + c (q_x^2+q_y^2),
\end{eqnarray}
with $q_{\pm} = q_x + i q_y$. The quadratic  energy dependence in the $q_xq_y$ plane stems from both the $f(\textbf{q})$ and $g(\textbf{q})$ functions, while the linear dispersion along $k_z$ appears only in $g(\textbf{q})$. Following analogous reasoning, it is straightforward to show that the momentum dependence of $d(\textbf{q})$ is identical to that of $g(\textbf{q})$, therefore
\begin{equation}
d(\textbf{q}) = m q_z + n (q_x^2+q_y^2).
\end{equation}
The energy dispersion around the band degeneracy is given by $E_{\pm}(\textbf{q})=d(\textbf{q})\pm \sqrt{f^2(\textbf{q})+g^2(\textbf{q})} = T(\textbf{k}) \pm U(\textbf{k})$. For elementary, linear Weyl nodes, this energy dispersion produces Fermi surfaces described by a quadric form with a singular point when chemical potential is located at the degeneracy. In this scenario, the type-I case with energy dispersion given by a conical surface producing closed Fermi surfaces appears when $T(\textbf{k}) < U(\textbf{k})$. The type-II case, with open Fermi surfaces, occurs when $T(\textbf{k}) > U(\textbf{k})$~\cite{Soluyanov2015}. This last condition implies that the cone is tilted. In the case of composite Weyl nodes, the division into type I and type II according to the relation between $T(\textbf{k})$ and $U(\textbf{k})$ terms still holds. However, the morphology of the Fermi surface is much more complicated due to the momentum dependence upto the  $4^{th}$ order, and especially due to the presence of two different tilting terms $m q_z$ and $n (q_x^2+q_y^2)$, that can result in a much richer variety of crossing types.

In our case, the velocities of the two bands along the $z$ direction are both positive, thus indicating the tilted scenario with the $m q_z$ term of the Hamiltonian being dominant. Thus, bcc iron presents a type-II composite Weyl node between bands 9 and 10. Volovik pointed out that a type-II Weyl node involves a Lifshitz transition between two Fermi surfaces at the Weyl node energy and the exchange of the Chern numbers of these surfaces~\cite{Volovik2017}. In bcc iron, Fermi surface sheets $S_{10,7}$ and $S_9$ present such behavior as shown in Ref.~\onlinecite{bccFe}.

\paragraph*{Nodal line.} The third type of band degeneracy is shown in Figure~\ref{fig:bands}(b). This degeneracy located at $k_y \approx 0.44 (2\pi/a)$ corresponds to a single point in a locus of points forming a nodal line in the $k_z=0$ mirror symmetry plane. In this plane, since the mirror symmetry $M_z$ operator and the Hamiltonian commute, the eigenstates of the Hamiltonian can be labeled by the $\pm i$ eigenvalues of the mirror symmetry operator. Therefore, bands with different labels may cross each other along lines in the symmetry plane. Nodal lines do not present a chiral charge and show a certain dispersion in energy. At the Fermi energy, the nodal lines pierce the Fermi surface at points where two Fermi surface sheets intersect. This behavior is illustrated in Figure~\ref{fig:WP}(c), where we show a small portion of the energy dispersion of bands 9 (red) and 10 (blue) in the $k_z=0$ plane around the $S_{10,4}$ electron pocket. One can see that the bands are degenerate along a line (black line) that connects two Fermi contours (red and blue lines). 

In summary, we have shown that a metal as simple as ferromagnetic bcc iron hosts a variety of band degeneracies, from elementary Weyl points to the new type-II composite Weyl points. These two chiral degeneracies play an essential role assigning bcc iron its topological character. We will now demonstrate the possibility of observing topological surface resonances in metals even when numerous bands cross the Fermi level.  

\section{\label{sec:surface_states} Surface states and resonances}

Fermi arc surface states that connect pairs of Weyl points of opposite chirality appear in the surface Brillouin zone if these two points are not projected onto each other. When the Fermi level is not precisely at the Weyl point energy, the arc-like surface states instead connect projected Fermi surface sheets characterized by the opposite Chern numbers \cite{PhysRevB.89.235315, haldane1}. The Fermi arcs emerge tangentially from the Fermi surfaces since the velocity of the surface states has to match the one of the projected bulk states \cite{haldane1}. In general, these surface states can overlap with the projection of the bulk states present at the Fermi level. In this situation, the surface state becomes a surface resonance hybridized with the bulk states. Indeed, this is an inherent difficulty encountered when identifying such surface states since the exact locations of the star and end points of the Fermi arc are obscured by the projections of bulk states.

As discussed above, bcc Fe is a topological metal that has two disconnected Fermi surface sheets with Chern numbers  $\mathcal{C}=\pm1$. However, due to its metallic character, the sBZ is almost entirely filled by the projected bulk states. One would naturally ask a question whether it is still possible to observe a surface state, or most likely, a surface resonance whose origin can be traced back to the presence of these two nontrivial electron pockets.

To address this question, we investigated surfaces states at the (100) and (110) surfaces, paying particular attention to the regions where the nontrivial electron pockets are projected. While at the (110) surface we were able to identify the small electron pocket among the other projected bulk states, on the (100) surface, the bulk states do not allow us to distinguish the $S_{10,6}$ and $S_{10,7}$ electron pockets. Therefore, below we will address only the (110) surface. 

\begin{figure*}
\includegraphics[width=.75\textwidth]{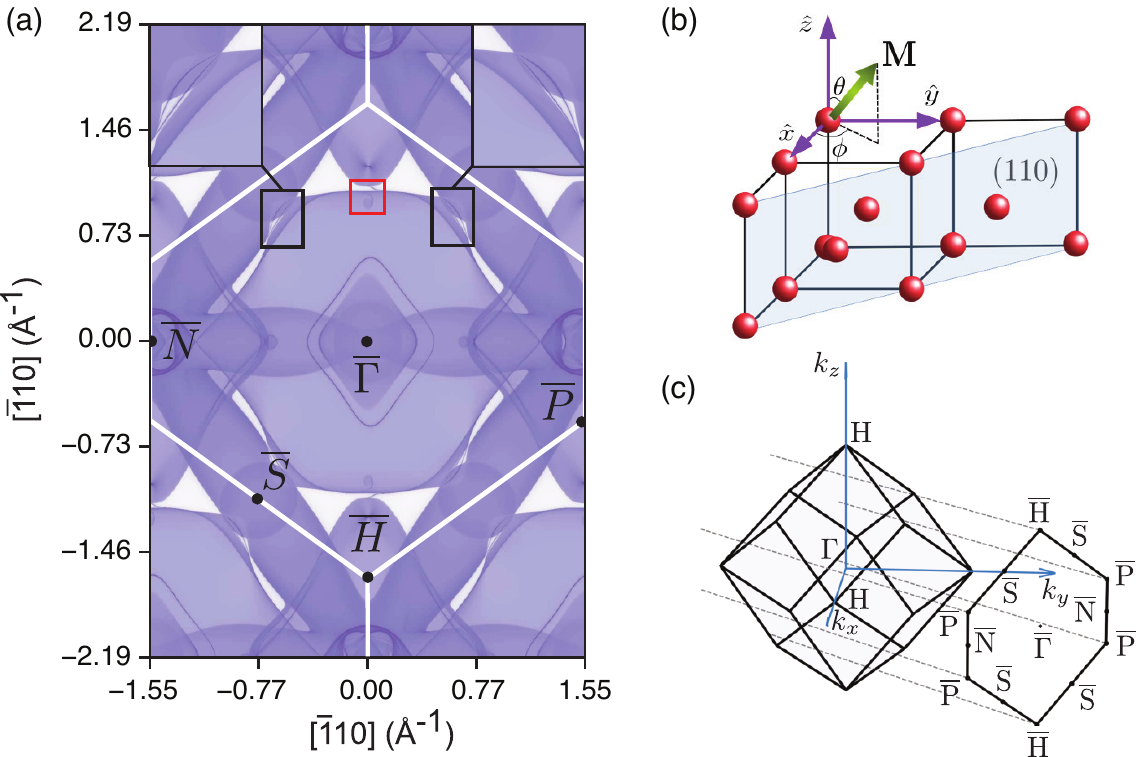} 
\caption{\label{fig:2} (a) Surface density of states calculated for the (110) surface of bcc iron. (b) Atomic structure of the (110) surface and the definition of the magnetization vector. (c) Relation between the bulk and surface Brillouin zones.}
\end{figure*}

\subsection{\label{subsec:surface_states:1} Surface density of states}

The surface density of states (DOS) of the (110) surface of bcc iron with the magnetization along the [001] direction is shown in Figure~\ref{fig:2}(a). In this plot, it is possible to identify the projected bulk states from most of partially occupied bands according to the shapes of their Fermi surface sheets (see Fig.~\ref{fig:1}(b)). In addition, in Figure~\ref{fig:2}(b) we illustrate the (110) surface and define the orientation of the magnetization. The relation between the bulk BZ and sBZ is given in Figure~\ref{fig:2}(c).

A higher density of projected bulk states is centered around the $\overline{N}$ point of the sBZ as shown in Figure~\ref{fig:2}(a). According to the Fermi surface sheets in Figure~\ref{fig:1}(b), bands 5, 6, 7 and 8 are present at point $\overline{N}$, although the contribution of each band is difficult to resolve. In addition, there are three distinct features that can be associated with particular sheets of the Fermi surface. First, the tubular structure produced by band 8 ($S_8$) is projected around the perimeter of the sBZ  crossing the central part of the sBZ along the $\overline{N}-\overline{\Gamma}-\overline{N}$ path. Second, we can identify the projection of the bulk states of band 9 enclosed by $S_9$ centered around the $\overline{\Gamma}$ point. These states cover most of the sBZ nearly touching the tubular region from band 8 at different points close to the middle of the $\overline{P}\overline{H}$ lines. Third, we observe the contribution from the bulk states of band 10 that correspond to the diamond-shaped region centered at $\overline{\Gamma}$, and four small nearly circular features in the middle of the $\overline{\Gamma}\overline{H}$ and $\overline{\Gamma}\overline{N}$ lines. These features correspond to the contributions originating from pockets $S_{10,1}$ and $S_{10,i}$ ($i=2,...,7$), respectively. The projections of the small electron pockets are divided into two groups, the four $S_{10,i}$ $(i=2,...,5)$ trivial electron pockets are projected in pairs onto the same location the sBZ along the $\overline{N}- \overline{\Gamma}- \overline{N}$ line, while the two nontrivial $S_{10,6}$ and $S_{10,7}$ electron pockets are projected separately onto the $ \overline{H} -\overline{\Gamma}- \overline{H}$ line.

\begin{figure}[t]
\includegraphics[width=.38\textwidth]{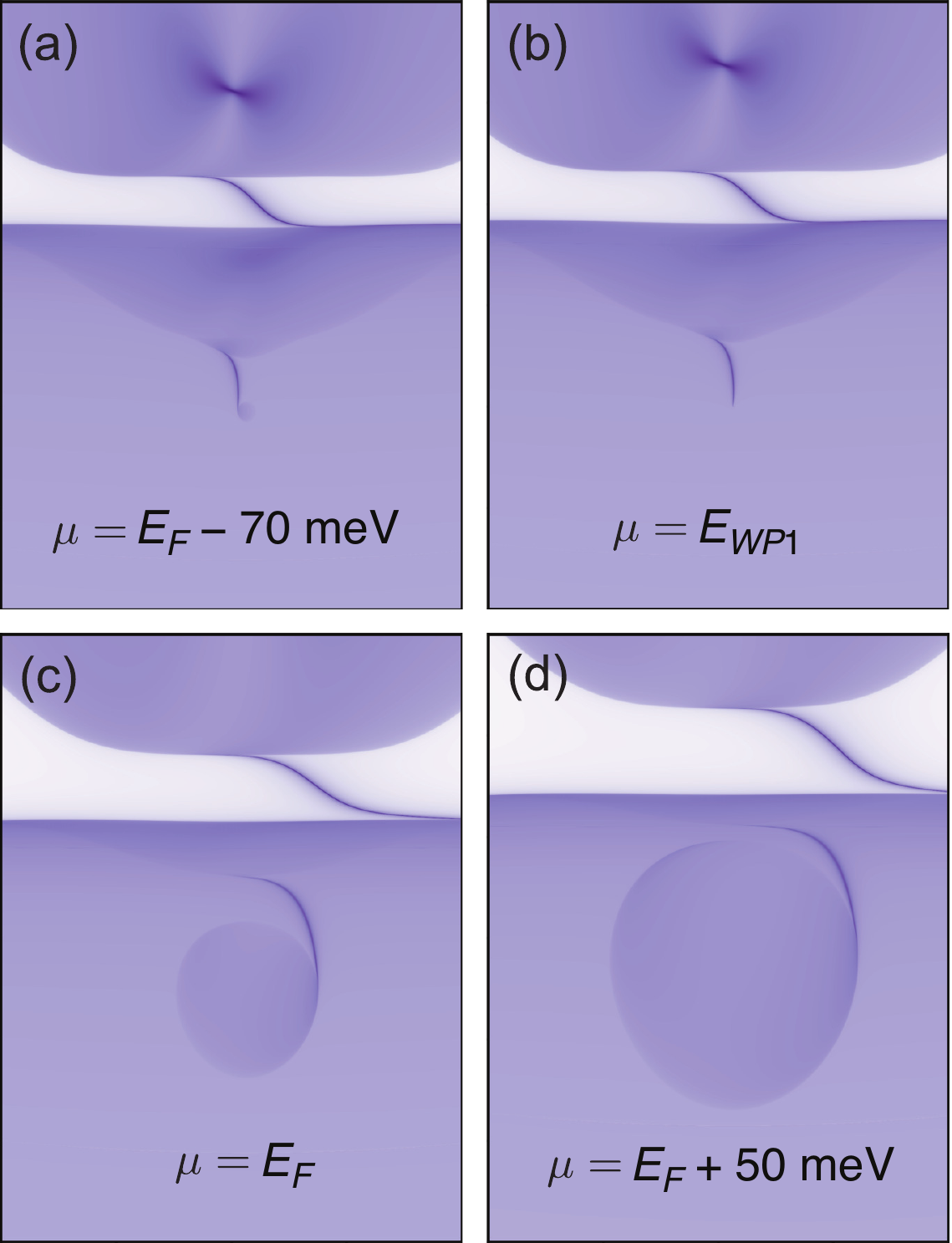} 
\caption{\label{fig:3} Evolution of the surface resonance related to pockets $S_{10,6}$ and $S_{10,7}$ upon changing chemical potential $\mu$. 
}
\end{figure}

Next, we focus on the features that correspond to surface states or surface resonances observed in Figure~\ref{fig:2}(a). In the area around the $\overline{N}$ points, the contribution of the projected bulk states to the surface density of states is high, and therefore, it is difficult to distinguish possible surface resonances. However, in the rest of the sBZ there are three different regions where surface states or resonances can be identified. First, we notice the presence of a diamond-shaped surface resonance that encircles the $\overline{\Gamma}$ point and the projection of the $S_{10,1}$ Fermi surface pocket. Since this surface resonance forms a closed loop, we conclude its topologically trivial origin. Furthermore, we found that this state is susceptible to the details of wannierization process, especially to the minimization of the spread function of the Wannier functions, which points to its possible relation to the surface dangling bonds. Second, in the regions indicated with black squares in Figure~\ref{fig:2}(a), where the projected bulk states of bands 8 and 9 nearly touch each other, appear two different surface states. These states behave differently in the left and the right parts of the sBZ (see insets in Figure~\ref{fig:2}(a)). The surface states on the left side of the sBZ emerge from the tubular surface $S_8$ and immerse again into $S_8$, while the surface states that appear on the right side of the sBZ emerge from the $S_9$ Fermi surface sheet and connect back to $S_9$. Since these surface states emerge from the projected bulk states, it is difficult to related their origin to the presence of any Weyl point, thus not allowing to argue about their nature. Interestingly, the existence of a surface resonance close to the Fermi level located at about three quarters of the $\overline{\Gamma}\overline{S}$ line has been observed experimentally~\cite{KIM2001193}. Third, the most interesting region is between the projections of pockets $S_{10,6}$ and $S_{10,7}$, indicated with a red square in Figure~\ref{fig:2}(a) and shown in detail in Figure~\ref{fig:3}(c). One can observe, on one hand, a short surface state that emerges from the projection of the bulk states of bands 8 and 9. Again, since this surface state emerges from two trivial $S_8$ and $S_9$ surfaces, it is difficult to argue about its nature and possible relation to any band degeneracy. On the other hand, we can also notice a surface resonance emerging from the projection of the nontrivial pockets $S_{10,6}$ and $S_{10,7}$. Below, we will demonstrate that this surface resonance originates from the chiral degeneracies in bcc iron. 

\subsection{\label{subsec:surface_states:2} Resonances at electron pockets $S_{10,6}$ and $S_{10,7}$}

In Figure \ref{fig:3} we plot the surface DOS at different values of chemical potential $\mu$ in the region of the sBZ where the states originating the $S_{10,7}$ pocket are projected. 
The projected bulk states correspond to three different bands. The lower part of each panel in Figure~\ref{fig:3} is occupied by the states of band 9 corresponding to the $S_9$ sheet, while in the upper part bulk states of band 8 are present. In the middle of the plots, it is possible to observe the projected states from the small electron pocket $S_{10,7}$. The electron pocket and the projected states of band 9 are connected through a resonance that emerges from both surfaces tangentially. Below, we will prove that this resonance originates from a band degeneracy enclosed by the $S_{10,7}$ sheet. It is also possible to observe a surface state that connects the projections of band 9 and 8. As explained previously, this surface state cannot be unambiguously associated to any chiral degeneracy, therefore, we will focus our attention only on the small resonance that connects the projected states of bands 9 and 10. 

The first evidence of the topological origin of this surface resonance is revealed by the manner it is attached to the projected bulk states of the $S_9$ and $S_{10,7}$ sheets. As pointed out by Haldane \cite{haldane1}, the contact to the projected surfaces is tangential, so the Fermi velocities of the surface and bulk states are equal. However, the situation considered here is slightly different to the simple scenario with only two Weyl nodes and a Fermi arc connecting the two nontrivial Fermi surfaces. In the discussed case, the surface resonance connects a nontrivial surface with a trivial one. One possible scenario to explain this discrepancy could be that the resonance hybridizes with the projected bulk states on the trivial surface, and re-emerges on the opposite side of the sBZ to connect with the complementary nontrivial electron pocket $S_{10,6}$.

A surface state that gives rise to theFermi arcs can have a complex energy dispersion, but it must follow the evolution of the Fermi surface and intersect the Weyl points. Accordingly, the surface resonance that emerges from $S_{10,7}$ should follow this trend if it has topological origin. In order to address the evolution of the surface resonance across the degeneracy energy, we have computed the surface DOS for different values of chemical potential $\mu$ close to the Fermi level $E_F$. Figure~\ref{fig:3} shows the surface DOS for chemical potentials at $E_F-70$~meV, the exact energy of the $WP1$ Weyl point $E_{WP1} = E_F - 63$~meV, $E_F$ and $E_F+50$~meV (panels a--d, respectively). In  Figure~\ref{fig:3}(a) we observe that the surface resonance emerges from the left side of a small hole pocket. Note that at $E_F-70$~meV the chemical potential is below $E_{WP1}$, thus the Fermi surface pocket $S_{10,7}$ corresponds to hole pocket. Upon increasing $\mu$ to $E_{WP1}$, the Fermi surface $S_{10,7}$ shrinks into a single point in the sBZ from which the resonance emerges (Fig.~\ref{fig:3}(b)). Finally, at higher energies the electron pocket extends, and the resonance emerges tangentially from the right side of the projected $S_{10,7}$ pocket (Figs.~\ref{fig:3}(c) and ~\ref{fig:3}(d)). In all these cases the resonance terminates in surface $S_9$ of the projected states of band 9. This behavior indicates that the resonance follows the expected behavior of a Fermi arc, especially at $\mu = E_{WP1}$ (Fig.~\ref{fig:3}(b)) where it emerges from the projection of the $WP1$ Weyl point.

\begin{figure}[t]
\includegraphics[width=0.35\textwidth]{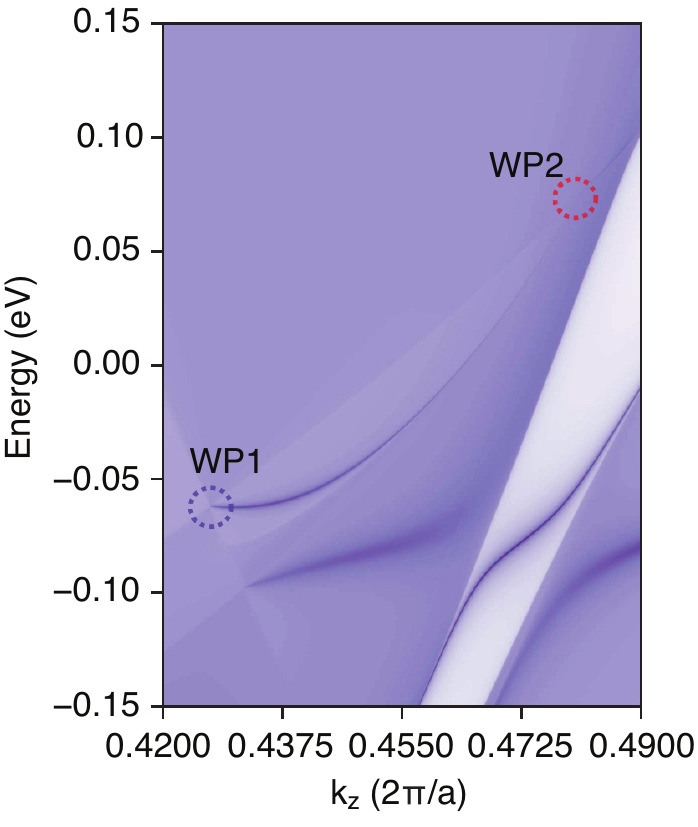}
\caption{\label{fig:5} Energy dispersion of the surface density of states calculated for the (110) surface of bcc iron along the $k_z$ direction.}
\end{figure}

Alternatively, the observed surface resonance can be associated to the  $WP1$ Weyl point by studying the projected band structure. In Figure~\ref{fig:5} we show the surface DOS along the $k_z$ direction in the region close to Weyl points $WP1$ and $WP2$. It is possible to observe a difference in contrast in those regions where the number of projected bands change. In particular, one can distinguish the contributions of the bands that give rise to Weyl points $WP1$ and $WP2$ as well as
a surface resonance crossing the Fermi level that emerges from $WP1$. This surface resonance disperses towards higher energies and merges with the projected band structure close to Weyl point $WP2$. Such energy dispersion suggests that the surface resonance connects Weyl nodes $WP1$ and $WP2$ rather than the two mirror-symmetric Weyl nodes  $WP1$. In addition to this surface resonance, one can distinguish other surface features: a surface resonance that emerges from the Weyl point between bands 8 and 9, suggesting that band degeneracies are a common cause of surface resonances in metallic systems, and a surface state that connects band 8 with band 9. 

Since Weyl nodes act as either sources or sinks of Berry curvature, a perturbation can only change their positions in momentum space, opening a gap only if Weyl nodes of opposite chiral charges annihilate each other. Thus, an external perturbation can either modify the topological state of a metal by changing the number, positions and types of chiral charges while leaving the system metallic or lead to a metal-insulator transition in which the resulting insulating phase can be both trivial and topological. In the first scenario, the Fermi surface Chern numbers will change since Weyl points can enter or leave the Fermi surface sheets through a Lifshitz transition, and the surface states or resonances will change accordingly. In the case of a metal-insulator transition resulting in a topological insulator state, the Fermi arcs have to transform into the topological surface states with a Dirac-cone band dispersion. As a matter of fact, Weyl semimetals were first proposed as an intermediate state between the trivial and topological insulators without inversion symmetry \cite{weyl_murakami}. 

\section{\label{sec:topo_tran} Topological phase transition}

\begin{figure*}[ht]
\centering
\includegraphics[width=0.9\textwidth]{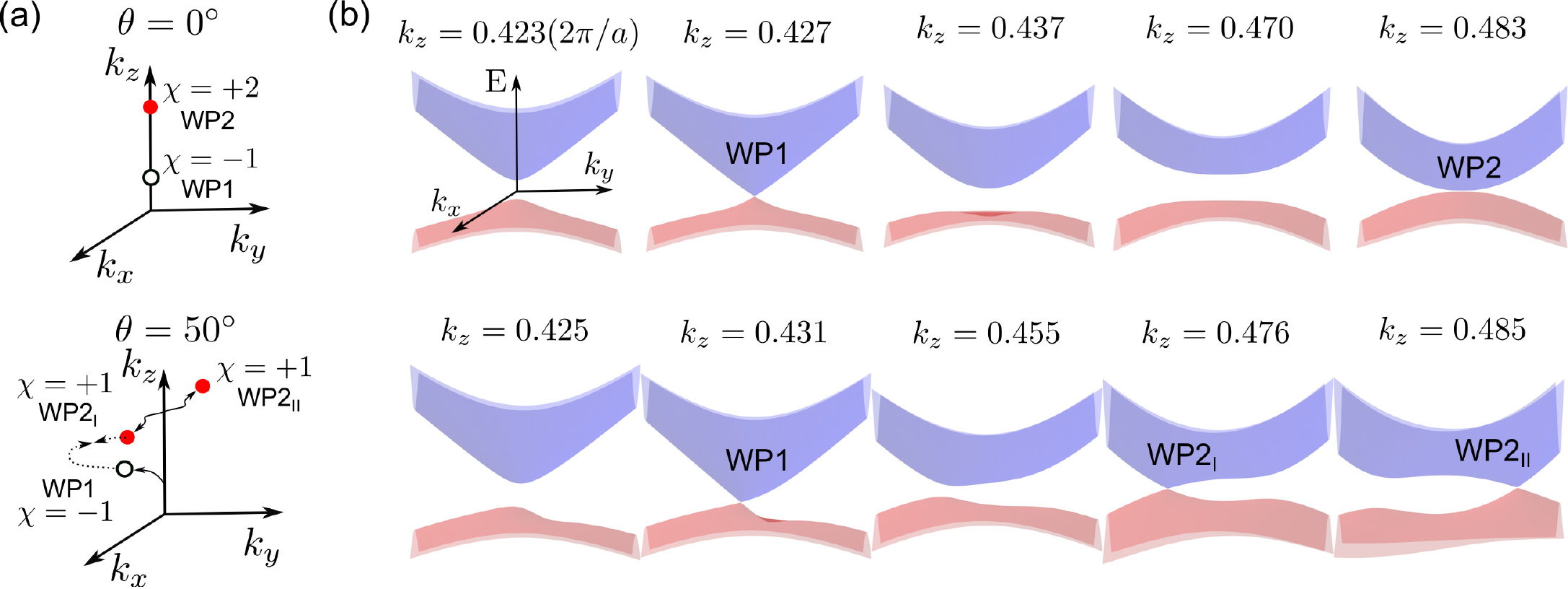}
\caption{\label{fig:wp_evolution} (a) Schematic illustration of the evolution of  Weyl points $WP1$ and $WP2$ upon changing the magnetization direction.
(b) Energy dispersion of bands 9 and 10 in the $k_x k_y$ plane at different values of $k_z$ for $\theta = 0^\circ$ and  $\theta = 50^{\circ}$, $\phi = 50^{\circ}$.
}
\end{figure*}

The possibility of manipulating and switching between topological phases by means of external parameters is an appealing idea from the point of view of potential applications. For example, the electronic structure of ferromagnetic systems can be easily controlled by changing the orientation of the magnetization by external magnetic field due to the spin-orbit coupling. This effect has been measured in bcc iron by M\l{}y\'nczak \emph{et al.} \cite{PhysRevX.6.041048}. Here, we will show that a topological phase transition in bcc Fe can be induced by manipulating adiabatically the orientation of the magnetization. 

\subsection{\label{subsec:topo_tran:1} Evolution of band degeneracies upon changing the magnetization direction}

When the magnetization of a bcc ferromagnet points along an arbitrary direction, there are no symmetry operations present \cite{PhysRev.172.498}. In this situation, the nodal lines and the composite Weyl nodes in bcc iron are no longer protected by the mirror and $C_4$ rotation symmetries, respectively. The nodal lines transform into individual Weyl points when the magnetization is tilted with respect to the easy axis $[001]$ \citep{bccFe}. Likewise, the two double Weyl nodes $WP2$ that are close to nontrivial electron pockets $S_{10,6}$ and $S_{10,7}$ split into pairs of elementary Weyl points ($WP2_{I}$ and $WP2_{II}$). Due to the proximity in $k$-space of the two points $WP1$ and $WP2$, we will have to follow the paths of all three Weyl nodes (the $WP1$ nodes and the two elementary Weyl nodes that compose $WP2$) upon adiabatic change of the magnetization vector.

We define the magnetization vector by the polar and azimuthal angles $\theta$ and $\phi$ in spherical coordinates (see Fig.~\ref{fig:2}(b)). We have analyzed the positions of the Weyl points at ten equidistant values of both the polar angle $\theta \in [0^{\circ},90^{\circ}]$ and the azimuthal angle $\phi \in [0^{\circ},90^{\circ}]$. In all the cases the double Weyl point $WP2$ splits into two elementary Weyl points $WP2_{I}$ and $WP2_{II}$ upon tilting the magnetization vector. As $\theta$ changes from $0^\circ$ to $90^\circ$, the $WP1$ point moves to larger $k_z$ values while the $WP2_{I}$ and $WP2_{II}$ points move away from each other. The $WP2_{II}$ point displaces to higher values of $k_z$ and $WP2_{I}$ tends to move to smaller values of $k_z$ approaching the $WP1$ Weyl point. 
This process is illustrated in Figure~\ref{fig:wp_evolution}(a) for the initial $\theta = 0^{\circ}$ and an intermediate value $\theta = 50^{\circ}$. In Figure~\ref{fig:wp_evolution}(b) we plot the energy dispersion in the $k_x k_y$ plane around the $(k_x,k_y)=(0,0)$ point and at different values of $k_z$. For the magnetization vector oriented along the easy axis $[001]$, i.e. $\theta=0^\circ$, bands 9 (red) and 10 (blue) touch only at points $k_z = 0.427 \pi/a$ and $k_z = 0.483 (2\pi/a)$, which correspond to Weyl nodes $WP1$ and $WP2$, respectively. The linear dispersion of the elementary Weyl node $WP1$ and the quadratic dispersion of the composite Weyl node $WP2$ are clearly seen. Upon tilting the magnetization vector to $\theta=50^{\circ}$ (in this case $\phi=50^{\circ}$), the $WP1$ degeneracy shifts to the left of the origin at $(k_x,k_y)=(0,0)$ while the $k_z$ value increases slightly to $k_z=0.431 (2\pi/a)$. The $WP2_{I}$ and $WP2_{II}$ degeneracies are observed at $k_z=0.476 (2\pi/a)$ and $k_z=0.485 (2\pi/a)$, respectively, both showing a linear band dispersion. The $WP2_{I}$ point displaced to the left of the origin at $(k_x,k_y)=(0,0)$, while $WP2_{II}$  moved in the opposite direction and towards higher values of $k_z$. 

\begin{figure}[b]
\centering
\includegraphics[width=0.38 \textwidth]{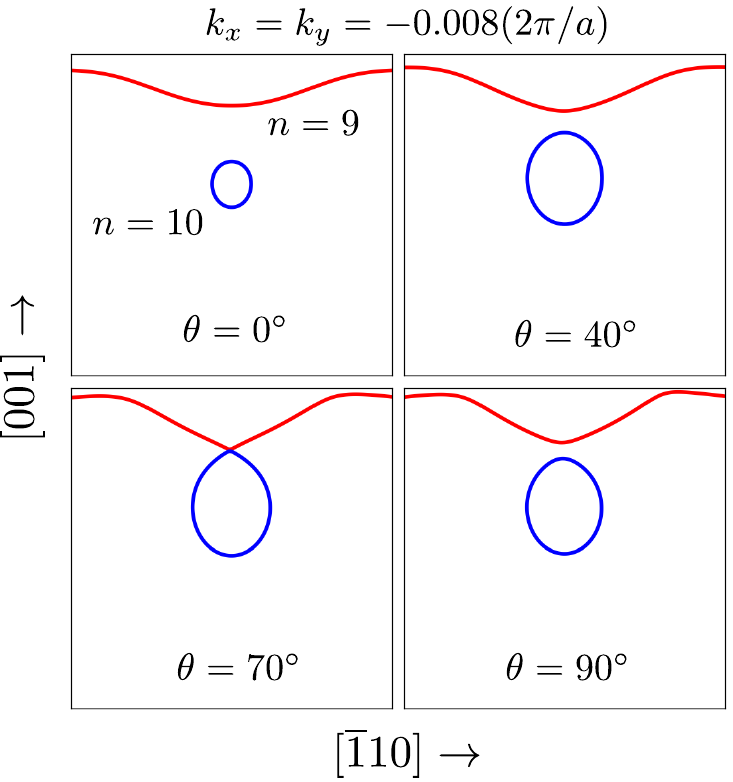}
\caption{\label{fig:lifshitz} Illustration of the Lifshitz transition involving Fermi surface sheets of bands 9 and 10 driven by the change of the magnetization direction.}
\end{figure}

\begin{figure*}[t]
\centering
\includegraphics[width=0.9\textwidth]{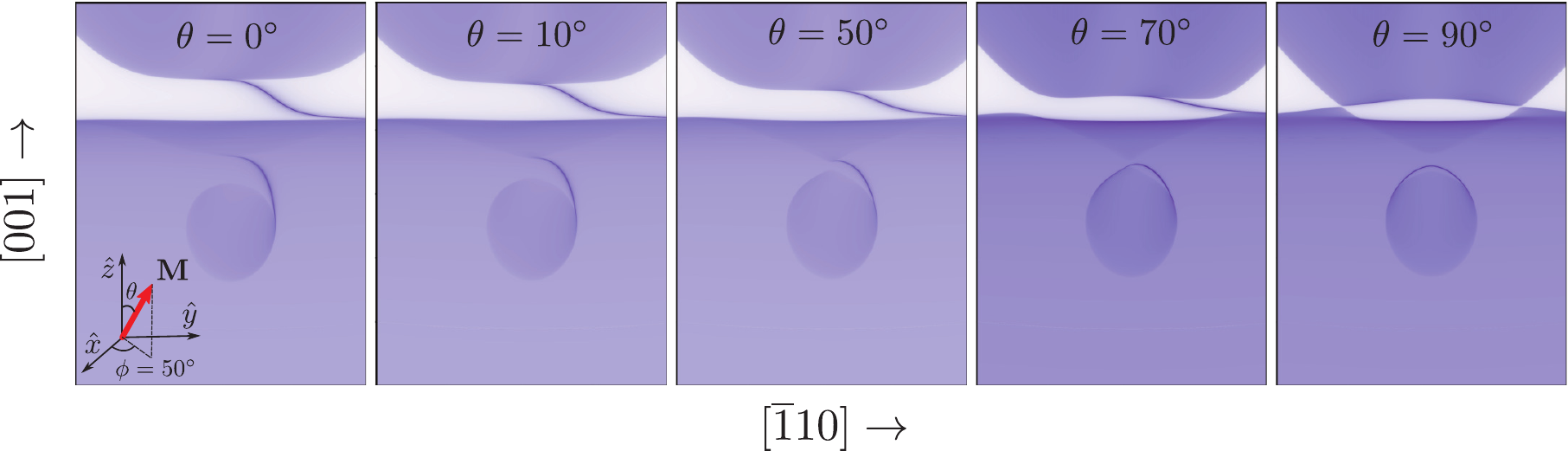}
\caption{\label{fig:L_transition} Evolution of the surface density of states around electron pocket $S_{10,7}$ upon changing $\theta$ at a fixed $\phi=50^{\circ}$.}
\end{figure*}

As Weyl points $WP1$ and $WP2_{I}$ with opposite chiral charges approach each other upon increasing $\theta$, these degeneracies eventually meet and annihilate. This general behavior is similar for different azimuthal angles $\phi$, although the exact value of polar angle $\theta$ at which the Weyl point annihilation takes place varies with $\phi$. Before the annihilation of the two Weyl points, $WP2_{I}$ has to enter the nontrivial electron pocket $S_{10,7}$. This process is realized through a Lifshitz transition between the two Fermi surface sheets $S_{10,7}$ and $S_9$ since $WP2_{I}$ is a type-II Weyl node tilted toward the $k_z$ axis akin to the original double node $WP2$. At the Lifshitz transition the two Fermi surface sheets exchange Chern numbers with the net value of the chiral charge of $WP2_{I}$. Note that the effective change of the Fermi surface Chern number of $S_9$ is zero because it is compensated by the exchange with the other nontrivial Fermi surface sheet $S_{10,6}$. After the transition, the topological state of bcc iron has changed since the net Chern number of $S_{10,7}$ is now zero because this Fermi surface sheet no longer contains any chiral degeneracy. 

Figure~\ref{fig:lifshitz} illustrates the Lifshitz transition upon changing $\theta$ for a fixed value $\phi=50^{\circ}$. We plot the Fermi contours by slicing along the $(001)$ and $(\overline{1}10)$ directions at $k_x=k_y=-0.008 (2\pi/a)$. The Fermi contours created by bands 9 (red) and 10 (blue) are initially separated at $\theta=0^{\circ}$. As the value of $\theta$ increases, the Fermi contour of band 10 expands and eventually touches that of band 9 at $\theta \approx 69^{\circ}$. The two Fermi contours detach upon further increase of $\theta$. According to our calculations, the two Weyl points $WP1$ and $WP2_{I}$ annihilate each other shortly after the Lifshitz transition. 

\subsection{\label{subsec:topo_tran:2} Evolution of the surface resonance upon changing the magnetization direction}

Finally, we address the evolution of the surface resonances that emerge from the nontrivial electron pockets $S_{10,6}$ and $S_{10,7}$ across the topological phase transition as a final confirmation of their topological origin. The change of the orientation of magnetization would not eliminate the surface resonances as long as the Fermi surface Chern numbers of pockets $S_{10,6}$ and $S_{10,7}$ do not change. A qualitative change of the surface resonances would be observed only following the discussed topological phase transition.

Figure~\ref{fig:L_transition} shows the surface DOS in the region of the sBZ around the projection of the $S_{10,7}$ pocket for different angles $\theta$ at a constant $\phi=50^\circ$. At $\theta < 65^{\circ}$ the surface resonance connects the projections of the $S_{10,7}$ and $S_9$ Fermi surface sheets. As $\theta$ increases, the projections of these two Fermi surface sheets approach each other, and the surface resonance shrinks accordingly. At the Lifshitz transition point, the projections of both surfaces are linked at the projection of the $WP2_{I}$ Weyl point, and the surface resonance connects the $S_{10,7}$ pocket projection with that of $WP2_{I}$. This is another indication that the $WP1$ and $WP2_{I}$ points are linked by a Fermi arc that gives rise to the surface resonance. At $\theta > 70^{\circ}$ the projections of the two Fermi surface sheets disconnect from each other, while the surface resonance persists as a trivial state that starts and ends in the projection of the now trivial $S_{10,7}$ sheet. 

The evolution of the shape and connectivity of the surface resonance can be understood in terms of the criteria for detecting Fermi arcs given in Refs.~\onlinecite{PhysRevX.5.031013,PhysRevLett.116.066802}. These criteria consist in counting the number of times surface states intersect a closed path at constant energy in the sBZ. A surface state counts as $+1$ or $-1$ depending on whether its Fermi velocity is along the direction of the path or opposite to it. If the path crosses an odd number of surface states and does not intersect any projection of bulk states, it encloses a nontrivial projected Fermi surface sheet. Otherwise, the enclosed Fermi surface sheet is trivial. In order to apply this rule to our case, we disregard the projected bulk states of band 9. In this case, for magnetization vectors characterized by $\theta < 69^{\circ}$ we can find a path that encloses the projected $S_{10,7}$ sheet and intersects the surface states only once. Past the topological phase transition for $\theta > 69^{\circ}$, the surface state is attached to $S_{10,7}$ from two sides, so we cannot find a path that does not cross the projection of $S_{10,7}$ and intersects the surface states an odd number of times. \\

\section{\label{sec:conclusion} Summary}

We addressed the possibility of observing surface resonances originating from the Weyl point band degeneracies and nontrivial Fermi surfaces in a well-studied prototypical ferromagnet -- bcc iron. We have shown that bcc iron can host a surface resonance at the (110) surface located along the $\overline{\Gamma} \overline{H}$ direction. This surface resonance is related to the Fermi arc that connects two Weyl points of different types close to the Fermi level. We further demonstrated that it is possible to manipulate this Fermi-arc surface resonance inducing a topological phase transition driven by the change of the magnetization orientation. 
Our study thus shows that Fermi-arc surface features can be observed in materials extending far beyond Weyl semimetals, and are possibly very common in a broad range of polar and magnetic compounds. Our work also establishes the methodology for identifying Fermi-arc surface states and resonances, proving their topological origin and designing control protocols.

%\appendix
%\section{}

\begin{acknowledgments}
We acknowledge support by the NCCR Marvel.  All first-principles calculations were performed at the Swiss National Supercomputing Centre (CSCS) under the projects s675 and s832.
\end{acknowledgments}

\bibliography{aps-bibliography}

\end{document}